
\magnification=\magstep 2

\overfullrule=0pt

\hfuzz=16pt

\baselineskip 20pt

\nopagenumbers\pageno=1\footline={\hfil -- {\folio} -- \hfil}

\

\centerline{{\bf Measurement of a Quantum System Coupled to}}
\centerline{{\bf Independent Heat-Bath and Pointer Modes}}

\

\centerline{Dima Mozyrsky and Vladimir Privman}

\ 
 
\centerline{\sl Department of Physics, Clarkson University}
\centerline{\sl Potsdam, New York 13699--5820, USA}

\centerline{\sl Electronic mail:\ \ {\tt privman@clarkson.edu}}

\vfill

\noindent{}To be published in{\bf\ Modern Physics Letters B \/}(2000).

\eject

\centerline{\bf Abstract}

\ 

We present an exact derivation of a process 
in which a microscopic measured system interacts with heat-bath
and pointer modes of a measuring device, via a coupling involving
a general Hermitian operator $\Lambda$ of the system. 
In the limit of strong interaction 
with these modes, over a small 
time interval, we derive the exact effective many-body
density matrix of the measured system plus pointer. We then discuss
the interpretation of the dynamics considered as the first stage in
the process of quantum measurement, eventually involving the wave-function 
collapse due to interactions with ``the rest of the universe.'' 
We establish that the effective density matrix represents 
the required framework for the measured system and 
the pointer part of the
measuring device to evolve into a statistical mixture described by
direct-product states
such that the system is in each eigenstate of $\Lambda$ with the correct
quantum-mechanical probability, whereas the expectation values of 
pointer-space operators retain amplified information of the system's
eigenstate.

\vfill\eject

The problem of quantum measurement has fascinated scientists
for a long time [1,2]. It has been argued that a large
``bath'' is an essential ingredient of the measurement
process. Interaction with the bath, which might be a heat-bath 
in thermal equilibrium, causes decoherence which is 
needed to form a statistical mixture of eigenstates out of the
initially fully or partially coherent quantum state of
the measured system. An ``external'' bath (``the rest of the universe'')
may also play a role
in selection of those quantum states of the pointer
that manifest themselves in classical observations [2-7].
In this work we propose a model in which the pointer retains 
information on the measurement result because of its 
coupling to the measured system, without the need to couple it also
to the internal bath. The measured system is still coupled to the 
internal bath.

In an exactly solvable model of a quantum oscillator coupled
to a heat bath of oscillators, it has been shown [4] that the
reduced density matrix of the system, with the bath traced over,
decoheres, i.e., it looses its off-diagonal elements in the
eigenbasis of the interaction Hamiltonian. Recent work on
decoherence [8-11] has explored the latter 
effect for rather general cases, for bosonic (oscillator) and 
spin baths. Applications for various physical systems have been
reported [12-18].
Fermionic heat bath has also been used in the literature [19].

It is clear, however, that the full function of a large, multimode
measuring device, interacting with a small (microscopic)
quantum system, must be different from thermal 
equilibration or similar averaging effect. The device
must store and amplify
the measurement outcome information. 
In this work we propose a solvable
model that shows how this is accomplished. 

It must be stressed that 
for a complete description of the measurement process one needs 
to interpret the transfer of the information stored after the 
system-pointer and system-internal bath interaction to the 
macroscopic level [2]. Our attention here is no the process which 
corresponds to the first stage 
of the measurement, in which the pointer acquires amplified information by
entanglement with the state of the system. Thus we do not claim to resolve
the foundation-of-quantum mechanics issue of how that information is 
passed on to the classical world, involving the collapse of the wave functions 
of the system and each pointer mode. Indeed, it is unlikely that
the wave function collapse can be fully described within the 
quantum-mechanical description of the three subsystems involved. 
Presumably, it would require
consideration of an external bath with which the pointer and the 
internal bath interact. This problem is not presently solved [1-3], 
and we first sidestep it by assuming separation of time scales (see 
below). However, we later argue that our results provide useful hints on 
how to view the larger problem of quantum measurement.

We now identify the three quantum systems involved. 
First,
the measured system, $S$, is a microscopic system with the
Hamiltonian which will be also denoted by $S$.
Second, the measuring device must
have the ``bath'' or ``body'' part, $B$, containing many individual
modes. The $k$th mode will have the Hamiltonian $B_k$.
The bath part of the device is not observed, i.e., it can be
traced over.
Finally, the device must also have modes
that are not traced over. These modes constitute the pointer, $P$,
that amplifies the information obtained in the measurement
process and can later pass it on for further amplification
or directly to macroscopic (classical) systems.
The $m$th pointer mode has the Hamiltonian $P_m$. It is expected that
expectation values of some quantities in the pointer undergo 
a large change during the measurement process.

It turns out, a posteriori,  that the device modes involved in the 
measurement process can be
quite simple, and they need not interact with
each other. This assumption allows us to focus on the
evolution of the system $S$ and its effect on the pointer $P$.
However, it is the pointer's interaction with the external
bath (some external modes, ``the rest of the universe'')
that is presumed to select those quantum states of $P$
that manifest themselves classically. For now, let us avoid the
discussion of this matter, see [2-6], by assuming that
the added evolution of the pointer due to such external interactions
occurs on time scales larger than the measurement time, $t$. Similarly,
when we state that the internal bath modes can be ``traced over,'' we 
really mean that their interactions with the rest of the universe are 
such that these modes play no role in the wave-function-collapse stage 
of the measurement process.

Furthermore, 
the measurement process probes
the wavefunction of the measured system at the initial time, $t=0$,
rather than its time evolution under $S$ alone. It is ideally
instantaneous. In practice, it is faster than the time scales
associated with the dynamics under $S$. Such a process
can be obtained as the limit of a system in which very strong
interactions between $S$ and $B$, and also between $S$ 
and $P$, are switched on at $t=0$ and switched off at $t>0$,
with small time interval $t$. At later times, the pointer can 
interact with other, external systems to pass on
the result of the measurement.

Thus, we assume that the Hamiltonian of the system itself, $S$,
can be ignored in the process. The total Hamiltonian
of the system plus device will be taken 
as
$$ H=\sum_k B_k + \sum_m P_m + b \Lambda 
\sum_k L_k +  p \Lambda \sum_m C_m  \eqno(1) $$
Here $\Lambda$ is some Hermitian operator of the system
that couples to certain operators of the modes, $L_k$
and $C_m$. The parameters $b$ and $p$ are
introduced to measure the coupling strength for the
bath and pointer modes, respectively. They are assumed very large;
the ideal measurement process corresponds to $b,p\to \infty$.

We note that the modes of $P$ and $B$ can be similar. The only 
difference between the bath and pointer modes is in how they interact 
with the ``rest of the universe'': the bath is traced over (unobserved), 
whereas the pointer modes have their wave functions collapsed in a later 
step of the measurement process. Thus, we actually took the same coupling 
operator $\Lambda$ for the bath and  pointer. In fact, all the exact 
calculations reported
in this work can be also carried out for different coupling
operators $\Lambda_b$ and $\Lambda_p$, for the bath and pointer
modes, provided they commute, $[\Lambda_b,\Lambda_p]=0$, so that
they share a common set of eigenfunctions. The final wavefunction
of the measured system, after the measurement, is in this set. 
Analytical calculation can be even extended to
the case when the system's Hamiltonian $S$ is retained in (1),
provided all three operators, $S,\Lambda_b,\Lambda_p$, 
commute pairwise. The essential physical ingredients of the model
are captured by the simpler choice (1).

We will later specify all the operators in (1) as
the modes of the bosonic heat bath of Caldeira-Leggett
type [17,19-26]. For now, however, let us keep our discussion
general. We will assume that the system operator $\Lambda$ 
has nondegenerate, discrete spectrum of 
eigenstates:
$$ \Lambda | \lambda \rangle = \lambda | \lambda \rangle \eqno(2) $$
Some additional
assumptions on the spectrum of $\Lambda$ and $S$ will be 
encountered later. We also note that the requirement that the coupling parameters
$b$ and $p$ are large may in practice be satisfied because, at the time
of the measurement, the
system's Hamiltonian $S$ corresponds to slow or trivial dynamics.

Initially, at $t=0$,
the quantum systems ($S,B,P$) and their modes
are not correlated with each other.
We assume that $\rho$ is the initial density matrix of 
the measured system. The initial state of each bath and pointer
mode will be assumed thermalized, with $\beta=1/(kT)$ and
the density matrices
$$ \theta_k = {e^{-\beta B_k}\over{\rm Tr}_k \left(e^{-\beta B_k}\right)}
\qquad\qquad
\sigma_m = {e^{-\beta P_m}\over{\rm Tr}_m \left(e^{-\beta P_m}\right)} \eqno(3) $$
We cannot offer any fundamental physical reason for having the initial
bath and pointer mode states thermalized, especially for the pointer; 
this choice is made to allow exact solvability.

The 
density matrix of the system at the time $t$ 
is
$$ R=e^{-iHt/\hbar}\rho\left(\prod_k \theta_k\right)\left( 
\prod_m \sigma_m\right) e^{iHt/\hbar} \eqno(4) $$
The bath is not probed and it can be traced over.
The resulting reduced density matrix $r$ 
of the combined system $S+P$ will be
represented by its matrix elements in the eigenbasis
of $\Lambda$. These
quantities are each an operator in the space 
of $P$:
$$ r_{\lambda\lambda^\prime}=\langle 
\lambda|{\rm Tr}_B(R)|\lambda^\prime\rangle \eqno(5)$$

We now assume that operators in different spaces and of different
modes commute. Then one can show 
that
$$ r_{\lambda\lambda^\prime}=\rho_{\lambda\lambda^\prime} \left[ \prod_m
e^{-i t \left( P_m + p \lambda C_m \right)/\hbar} \sigma_m 
e^{i t \left( P_m + p \lambda^\prime C_m \right)/\hbar} \right] \times $$
$$ \left[ \prod_k {\rm Tr}_k \left\{ e^{-i t \left( B_k + b \lambda L_k 
\right)/\hbar} \theta_k 
e^{i t \left( B_k + b \lambda^\prime L_k \right)/\hbar} \right\} \right] \eqno(6) $$
where 
$\rho_{\lambda\lambda^\prime}=\langle 
\lambda| \rho |\lambda^\prime\rangle$.
This result involves products of $P$-space operators
and traces over $B$-space operators which are all single-mode. Therefore,
analytical calculations are possible for some choices of the
Hamiltonian (1). The observable $\Lambda$ can be kept general.

The role of the product of traces over the modes of the bath 
in (6) is to induce decoherence
which is recognized as essential for the 
measurement process, e.g., [1,2]. At the time $t$, the absolute
value of this product should approach $\delta_{\lambda\lambda^\prime}$
in the limit of large $b$. Let us now assume that the bath is 
bosonic. The Hamiltonian
of each mode is then $ \hbar\omega_k a^\dagger_k a_k$, where
for simplicity we shifted the zero of the oscillator energy to the
ground state. The coupling operator $L_k$ is usually selected as
$L_k=g^*_ka_k + g_k a_k^\dagger$. For simplicity, though,
we will assume that the coefficients $g_k$ are 
real:
$$B_k = \hbar \omega_k a^\dagger_k a_k
\qquad\quad\quad 
L_k=g_k\left(a_k + a_k^\dagger\right) \eqno(7)$$
For example, for radiation field in a unit volume,
coupled to an atom [27],
the coupling is via a linear combination of the operators
$(a_k + a_k^\dagger )/\sqrt{\omega_k}$ and 
$i(a_k - a_k^\dagger )/\sqrt{\omega_k}$. For a spatial oscillator,
these are proportional to position and momentum, respectively. 
Our calculations can be extended to have an
imaginary part of $g_k$ which adds interaction with momentum.

The product of traces in (6) can be calculated by coherent-state
or operator-identity techniques [8-10]. Here and below we only
list the results of such calculations which are usually quite 
cumbersome:
$$  \prod_k {\rm Tr}_k \{ ... \}  = \exp\left\{-2b^2
\left(\lambda-\lambda^\prime\right)^2 \Gamma (t)
+ib^2\left[\lambda^2-(\lambda^\prime)^2\right]\gamma(t) \right\}\eqno(8) 
$$
$$
 \Gamma(t)= \sum_k (\hbar\omega_k)^{-2} g_k^2
\sin^2 {\omega_k t\over 2}
\coth{{\hbar\beta\omega_k \over 2}} \eqno(9) $$
Explicit 
form of $\gamma (t) $ is also known [8].

In the continuum limit of many modes, the density of the bosonic 
bath states in unit volume,
$D(\omega)$, and the Debye cutoff with frequency, $\omega_D$,
are introduced [22] to 
get
$$ \Gamma(t)= \int\limits_0^\infty \, d\omega \,   
{D(\omega) g^2(\omega)\over(\hbar\omega)^2}\, 
e^{-\omega/\omega_D}\, \sin^2 {\omega t\over 2}
\coth{{\hbar\beta\omega \over 2}} \eqno(10) $$
Let us consider the popular choice 
termed Ohmic dissipation [22],
motivated by atomic-physics [27] and
solid-state applications [22], corresponding 
to
$$ D(\omega) g^2(\omega) = \Omega \, \omega \eqno(11)$$
where
$\Omega$ is a constant. Other powers of 
$\omega$ have also been 
considered, e.g., [11].
In studies of decoherence [8-11] for large times $t$, for models without
strong coupling,
not all the choices of $D(\omega) g^2(\omega)$ 
lead to complete decoherence [11] because $\Gamma (t)$
must actually diverge to $+\infty$ for
$t\gg \hbar \beta$, as happens for the choice (11). 

Let us assume
that the energy gaps of $S$ are bounded so that there
exists a well defined time scale $\hbar/\Delta S $ of the evolution
of the system under $S$. There is also
the time scale $1/\omega_D$ set by the frequency cutoff assumed
for the interactions. The thermal time scale is $\hbar \beta$. 
The only real limitation on the duration of measurement
is that $t$ must be less then $\hbar/\Delta S$. In applications,
typically [22] one can assume that $1/\omega_D \ll \hbar/\Delta S$.
Furthermore, it is customary to assume that the temperature is 
low [22], 
$$ t {\rm\ and\ } 1/\omega_D \ll \hbar/\Delta S \ll \hbar\beta  \eqno(12) $$
In 
the limit of large $\hbar\beta$, the absolute value of (8) reduces 
to
$$ {\rm Abs} \! \left( \prod_k {\rm Tr}_k \{ ... \} \right) \simeq
 \exp\left\{ -{\Omega
\over 2 \hbar^2} b^2
\left(\lambda-\lambda^\prime\right)^2 \ln [1+(\omega_D t)^2]
\right\}\eqno(13) $$
In 
order to achieve effective decoherence, the product 
$ (\Delta \lambda)^2\, b^2  \ln [1+(\omega_D t)^2] $
must be large. The present 
approach only applies to operators $\Lambda$
with nonzero scale of the smallest spectral gaps, $\Delta \lambda$.

We note that the decoherence property needed for
the measurement process will be obtained for nearly any well-behaved
choice of $D(\omega)g^2(\omega)$ because we can rely on the value
of $b$ being large rather than on the properties of
the function $\Gamma (t)$. If $b$ can be large enough,
very short measurement times are possible. However, it may be advisable
to use measurement times $ 1/\omega_D \ll t \ll \hbar/\Delta S$ to 
get the extra amplification factor $\sim \ln (\omega_D t)$ and allow
for fuller decoherence and less sensitivity to the value of $t$ in the
pointer part of the dynamics, to be addressed shortly.
We notice, furthermore, that the assumption
of a large number of modes is important for monotonic decay
of the absolute value of (8) in decoherence studies [8-11], where irreversibility
is obtained only in the limit of infinite number of modes. In our case, it
can be shown that 
such a continuum limit allows to 
extend the possible measurement times
from $t \ll 1/\omega_D$ to $1/\omega_D \ll t \ll \hbar/\Delta S$.

Consider the reduced density matrix
$r$ of $S+P$, see (6). It becomes diagonal in $|\lambda\rangle$,
at the time $t$, because all the nondiagonal elements are  
small,
$$ r=\sum_\lambda |\lambda \rangle \langle \lambda | \, \rho_{\lambda\lambda}
\prod_m e^{-i t \left( P_m + p \lambda C_m \right)/\hbar} \sigma_m 
e^{i t \left( P_m + p \lambda C_m \right)/\hbar} \eqno(14)$$
Thus, the described stage of the measurement process yields the density 
matrix that can be interpreted as describing a statistically distributed
system, without quantum correlations. This, however, is only meaningful within the 
ensemble interpretation of quantum mechanics. 

For a single system plus device, coupling to the rest of the universe is 
presumably needed (this problem is not fully understood in our opinion, 
see [2]) for that system to be left in one of the eigenstates $|\lambda \rangle$,
with probability $\rho_{\lambda\lambda}$. After the
measurement interaction is switched off at $t$, the pointer
coupled to that system will carry information on the value of $\lambda$. 
This information is ``amplified,''
owing to the large parameter $p$ in the interaction. 

We note that one of the roles of the pointer having many modes, many of which 
can be identical and noninteracting, is to allow it (the pointer only) to still be 
treated in the ensemble, density matrix description, even if we focus on the 
later stages of the measurement when the wave functions of a single measured 
system and of each pointer mode are already collapsed. This pointer density 
matrix can be read off (14). This aspect is new and it may provide a useful hint 
on how to set up the treatment of the full quantum-measurement process description. 

Another such hint is provided by the fact that, as will be shown shortly, 
the changes in the expectation values of some observables of the pointer 
retain amplified information on the system's eigenstate. So, coupling to the 
rest of the universe that leads to the completion of the measurement process, 
should involve such an observable of the pointer. Eventually, the information 
in the pointer, perhaps after several steps of amplification,
should be available for probe by interactions with classical devices. 

At time $t=0$, expectation values of various operators 
of the pointer will have their initial values. These values will
be different at the time $t$ of the measurement owing to the
interaction with the measured system. It is expected that
the large coupling parameter $p$ will yield large 
changes in expectation values of the pointer quantities. 
This does not apply equally to
all operators in the $P$-space. Let us begin with the simplest choice: 
the Hamiltonian $\sum\limits_m P_m$
of the pointer. 

We will assume that the pointer is described by
the bosonic heat bath and, for simplicity, use the same notation
for the pointer modes as that used for the bath modes. The assumption
that the pointer modes are initially thermalized, see (3),
was not used thus far.
While it allows exact analytical calculations, it is not essential: the
effective density matrix describing the pointer modes at the time $t$, for 
the system's state $\lambda$, will retain amplified information on the value 
of $\lambda$ for general initial states of the pointer. 

This effective density matrix is the product over the $P$-modes in (14). 
For the ``thermal'' $\sigma_m$ from (3), the
expectation value of the pointer energy $E_P$ can be calculated 
from
$$ \langle E_P \rangle_\lambda\,{\rm Tr}_P \left(e^{-\hbar\beta \sum_s 
\omega_s a^\dagger_s a_s }\right)=
{\rm Tr}_P \left\{ \left( \sum_m \hbar\omega_m a^\dagger_m a_m 
\right) \right. \times \qquad\qquad  \eqno(15)
$$
$$
\left. \prod_n \left[ e^{-i t [ \omega_n a^\dagger_n a_n + p \lambda 
g_n(a_n + a_n^\dagger) ]/\hbar} \left(e^{-\hbar\beta \sum_k 
\omega_k a^\dagger_k a_k}\right) e^{i t [ \omega_n a^\dagger_n a_n + p \lambda 
g_n(a_n + a_n^\dagger) ]/\hbar}\right] \right\} .$$
The right-hand side can be reduced to calculations for individual modes.
Operator identities can be then utilized to obtain the 
results
$$ \langle E_P \rangle_\lambda (t) = \langle E_P 
\rangle (0) + \langle \Delta E_P \rangle_\lambda (t)\eqno(16)
$$
$$ 
\langle E_P \rangle (0) = \hbar \sum_m \omega_m e^{-\hbar\beta \omega_m} \left(1-
e^{-\hbar\beta \omega_m}\right)^{-2} \eqno(17) 
$$
$$
\langle \Delta E_P \rangle_\lambda (t)=  {4 p^2 \lambda^2 \over \hbar} 
\sum_m {g_m^2 \over 
\omega_m} \sin^2 \left({\omega_m t \over 2}\right) \eqno(18) $$
For 
a model with Ohmic dissipation, the resulting integral, in the
continuum limit, can be calculated to 
yield
$$ \langle \Delta E_P \rangle_\lambda (t)=  {2\, 
\Omega \,\omega_D \lambda^2 p^2 \over \hbar}
{(\omega_Dt)^2 \over 1 + (\omega_Dt)^2} \eqno(19)$$
which
 should be compared to the exponent in (13). The energy will be an indicator
of the amplified value of the square of 
$\lambda$, provided $p$
is large. Furthermore, we see here the advantage of
larger measurement times, $t \gg 1/\omega_D$. The change in the
energy then reaches saturation. After the time $t$, when the
interaction is switched off, the energy of the pointer
will be conserved.

Let us consider the expectation value of the following 
Hermitian operator of the 
pointer:
$$ X=\sum_m C_m = \sum_m g_m (a_m +a_m^\dagger) \eqno(20) $$
For
an atom in a field, $X$ is related to the electromagnetic
field operators [24]. One can show that
$\langle X_P \rangle (0) =0$ 
and
$$ \langle \Delta X_P \rangle_\lambda (t)=  \langle 
X_P \rangle_\lambda (t) = -{4 p \lambda \over \hbar} \sum_m {g_m^2 \over 
\omega_m} \sin^2 \left({\omega_m t \over 2}\right) 
\qquad\qquad \eqno(21) 
$$
$$ 
=  -{2\, 
\Omega \,\omega_D \lambda p \over \hbar}
{(\omega_Dt)^2 \over 1 + (\omega_Dt)^2} $$
The 
change in the expectation value of $X$ is linear in $\lambda$.
However, this operator is not
conserved. One can show that after the time $t$ its expectation value 
decays to zero for times $t + {\cal O} (1/\omega_D)$.

We note that by referring to ``unit volume'' we have avoided
the discussion of the ``extensivity'' of various quantities. 
For example, the initial energy $\langle E_P 
\rangle (0)$ is obviously
proportional to the system volume, $V$. However, 
the change 
$\langle \Delta E_P \rangle_\lambda (t)$ 
will not be extensive; typically,
$g^2(\omega)\propto 1/V$, $D(\omega) \propto V$. Thus, while the 
amplification
in our measurement process can involve a numerically large 
factor, the changes in the quantities
of the pointer will be multiples of microscopic values. Multi-stage
amplification, or huge coupling parameter $p$, would be needed for
the information in the pointer to become truly ``extensive'' macroscopically.

In practice, there will be probably two types of pointer involved in a multistage
measurement process. Some pointers will consist of many 
{\it noninteracting\/} modes.
These pointers carry the information, stored in a density matrix rather
than a wave function of a single system. The latter transference hopefully makes the
wavefunction collapse and transfer of the stored information to the macroscopic level less
``mysterious and traumatic.'' The second type of pointer will involve 
strongly
interacting modes and play the role of an amplifier by utilizing the 
many-body
collective behavior of the coupled modes (phase-transition style). Its role will be
to alleviate the artificial requirement for large mode-to-system coupling
parameters encountered in our model.

In summary, we described the first stage of a measurement process. 
It involves decoherence
due to a bath and transfer of information 
to a large system (pointer) via
strong interaction over a short period of time.
The pointer itself need not be coupled to the internal bath.
While we do not offer a solution to the foundation-of-quantum-mechanics
wave-function collapse problem [2], our results do provide two interesting observations. 

Firstly, the pointer operator ``probed'' by the rest of the universe during 
the wave-function 
collapse stage, may be in part determined not only by how the pointer modes are 
coupled to the external bath [3-7], 
but also by the amplification capacity of that
operator in the first stage of the process, as illustrated by our calculations.

Secondly, 
for a single system (rather then an ensemble), the multiplicity of the 
(noninteracting) 
pointer modes might allow the pointer to be treated within the density 
matrix 
formalism even after the system and each pointer-mode wave functions were 
collapsed. Since it is the information in the pointer that is passed on,
this observation might seem to resolve part of the measurement puzzle.
Specifically, it might suggest why only those density matrices entering 
(14) are selected for the pointer: they carry classical (large, different 
from 
other values) information in expectation values, rather than quantum-mechanical 
superposition.  However, presumably [2] only a full description of the 
interaction of the external world with the system $S+P$ can explain the 
wavefunction collapse of $S$. 

We acknowledge helpful discussions with
Professor L.~S.~Schulman. This research has been supported
by the US Army Research Office under grant 
DAAD$\,$19-99-1-0342.

\vfil\eject

\centerline{\bf References}{\frenchspacing
 
\item{1.} For a historical overview, see, A. Whitaker, 
{\sl Einstein, Bohr and the Quantum Dilemma\/} (Cambridge 
Univ. Press, Cambridge, 1996).

\item{2.} J. Bell, Phys. World {\bf 3}, August 1990, No. 8, p. 33.

\item{3.} W. H. Zurek, Physics Today, October 1991, p. 36.

\item{4.} W. G. Unruh, W. H. Zurek, Phys. Rev. D {\bf 40},
1071 (1989).

\item{5.} W. H. Zurek, S. Habib and J. P. Paz,
Phys. Rev. Lett. {\bf 70}, 1187 (1993).

\item{6.} M. Gell-Mann and J. B. Hartle, in 
{\sl Proceedings of the 25th International Conference on High Energy 
Physics\/} (South East Asia Theor. Phys. 
Assoc., Phys. 
Soc. of Japan, Teaneck, NJ, 1991) Vol. 2, p. 1303. 

\item{7.} M. Gell-Mann and J. B. Hartle, in  {\sl Quantum 
Classical Correspondence: The 4th Drexel Symposium on Quantum 
Nonintegrability}, ed. by D. H. Feng and B. L. Hu
(International Press, Cambridge, MA, 1997) p. 3.

\item{8.} D. Mozyrsky and V. Privman, J. Stat. Phys. {\bf 91},
787 (1998).

\item{9.} N. G. van Kampen, J. Stat. Phys. {\bf 78},
299 (1995).

\item{10.} J. Shao, M.-L. Ge and H. Cheng, Phys. Rev. E {\bf 53},
1243 (1996).

\item{11.} G. M. Palma, K. A. Suominen and A. K. Ekert,
Proc. Royal Soc. London A {\bf 452}, 567 (1996).

\item{12.} I. S. Tupitsyn, N. V. Prokof'ev, P. C. E. Stamp,
Int. J. Modern Phys. B {\bf 11}, 2901 (1997).

\item{13.} C. W. Gardiner, {\sl Handbook of Stochastic Methods
for Physics, Chemistry and the Natural Sciences}
(Springer-Verlag, Berlin, 1990).

\item{14.} A. J. Leggett, in {\sl Percolation, Localization and
Superconductivity, NATO ASI Series B: Physics}, 
ed. by A. M. Goldman and S. A. Wolf (Plenum, NY, 1984), 
Vol. 109, p. 1.

\item{15.} J. P. Sethna, Phys. Rev. B {\bf 24}, 698 (1981).

\item{16.} Review: A. O. Caldeira and A. J. Leggett,
Ann. Phys. {\bf 149}, 374 (1983).

\item{17.} A. Garg, Phys. Rev. Lett. {\bf 77}, 764 (1996).

\item{18.} L. Mandel and E. Wolf, {\sl Optical Coherence 
and Quantum Optics} (Cambridge Univ. Press, Cambridge, 1995).

\item{19.} L.-D. Chang and S. Chakravarty, Phys. Rev. B
{\bf 31}, 154 (1985).

\item{20.} A. O. Caldeira and A. J. Leggett,
Phys. Rev. Lett. {\bf 46}, 211 (1981).

\item{21.} S. Chakravarty and A. J. Leggett, Phys. Rev. Lett.
{\bf 52}, 5 (1984).

\item{22.} Review: A. J. Leggett, S. Chakravarty, A. T. Dorsey,
M. P. A. Fisher and W. Zwerger, Rev. Mod. Phys. {\bf 59}, 1
(1987) [Erratum {\it ibid.\/} {\bf 67}, 725 (1995)].

\item{23.} A. O. Caldeira and A. J. Leggett, Physica {\bf 121A},
587 (1983).

\item{24.} R. P. Feynman and A. R. Hibbs,
{\sl Quantum Mechanics and Path Integrals} 
(McGraw-Hill Book Co., NY, 1965).

\item{25.} G. W. Ford, M. Kac and P. Mazur,
J. Math. Phys. {\bf 6}, 504 (1965).

\item{26.} A. J. Bray and M. A. Moore, Phys. Rev. Lett.
{\bf 49}, 1546 (1982).

\item{27.} W. H. Louisell, {\sl Quantum Statistical
Properties of Radiation} (Wiley, NY, 1973).

}
 
\bye